\documentclass[%
 reprint,
 amsmath,amssymb,
 aps,
]{revtex4-1}

\usepackage{graphicx}
\usepackage{dcolumn}
\usepackage{bm}
\usepackage{units}
\usepackage{float}

\newcommand{\bracket}[2]{\ensuremath{\left\langle #1 \middle| #2 \right\rangle}}
\newcommand\Tstrut{\rule{0pt}{2.6ex}}       
\newcommand\Bstrut{\rule[-1.1ex]{0pt}{0pt}} 

\begin{document}

\preprint{APS/123-QED}

\title{Single Particle Strengths and Mirror States in $^{15}$N$-^{15}$O 
below 12.0 MeV}%


\author{C.E. Mertin$^{1}$, D.D. Caussyn$^{1}$, A.M. Crisp$^{1}$, 
N. Keeley$^{2}$, \\K.W. Kemper$^{1}$, O. Momtyuk$^{1}$, B.T. Roeder$^{1}$, 
and A. Volya$^{1}$\vspace{.5em}}
\affiliation{%
 $^{1}$Department of Physics, Florida State University, Tallahassee, Fl, 32312\\
 $^{2}$National Centre for Nuclear Research, Andrzeja So\l tana 7, 05-400 
Otwock, Poland
}%


\begin{abstract}
New $^{14}$N(d,p) angular distribution data were taken at a deuteron 
bombarding energy of 16 MeV to locate all narrow single particle 
neutron states up to 15 MeV in excitation. A new shell model calculation 
is able to reproduce all levels in $^{15}$N up to 11.5 MeV and is used 
to characterize a narrow single particle level at 11.236 MeV and to 
provide a map of the single particle strengths. The known levels in 
$^{15}$N are then used to determine their mirrors in the lesser known 
nucleus $^{15}$O. The 2s$_{\unitfrac{1}{2}}$ and 1d$_{\unitfrac{5}{2}}$ 
single particle centroid energies are determined for the $^{15}$N$-^{15}$O 
mirror pair as:~$^{15}$N~$\left(\text{2s}_{\unitfrac{1}{2}}\right)=8.08$~MeV, 
$^{15}$O~$\left(\text{2s}_{\unitfrac{1}{2}}\right)=7.43$~MeV, 
$^{15}$N~$\left(\text{1d}_{\unitfrac{5}{2}}\right)=7.97$~MeV, and 
$^{15}$O~$\left(\text{1d}_{\unitfrac{5}{2}}\right)=7.47$~MeV. These results 
confirm the degeneracy of these orbits and that the $^{15}$N$-^{15}$O nuclei 
are where the transition between the 2s$_{\unitfrac{1}{2}}$ lying below the 
1d$_{\unitfrac{5}{2}}$ to lying above it, takes place. The 
1d$_{\unitfrac{3}{2}}$ single particle strength is estimated to be 
centered around 13 MeV in these nuclei.

\begin{description}
\item[PACS numbers]
21.60.Cs, 21.10.Jx, 25.45.Hi
\end{description}
\end{abstract}

\maketitle

\section{\label{sec:level1}Introduction}
 It was realized early in the development of theoretical nuclear structure 
models \cite{Hal57} that the $^{15}$N nucleus is an ideal candidate for study 
because considerable experimental work had shown that there existed seven 
positive parity states that might be described as a closed 1p shell with a 
single 2s or 1d shell particle outside its core. In addition, the large 
energy gap between the ground and first excited state, 5.3 MeV, similar to 
the 6 MeV gap in $^{16}$O, reinforced the idea of $^{15}$N having a closed 
core for its ground state. It was also pointed out in Halbert and French 
\cite{Hal57}, one of the first shell model calculations describing the low 
lying positive parity states, that many levels in $^{15}$N could be populated 
by a large variety of inelastic and particle transfer reactions making this 
nucleus ideal for testing details of future model calculations. Later, weak 
coupling model calculations \cite{Hsi70, Lie70} focused on the positive 
parity states with the goal of extending the understanding of these states 
up to 10 MeV in excitation. While the first works included only 1p-2h 
configurations it was argued by Shukla and Brown \cite{Shu68} that  
contributions from 3p-4h states were needed to describe several of the levels 
below 10 MeV and the work of Lie {\em et al.} \cite{Lie70} confirmed this 
idea. Lie {\em et al.} \cite{Lie70} also proposed that  spin parity 
assignments could be made by comparing theoretical calculations with the 
measured properties of levels and in further work Lie and Engeland 
\cite{Lie76} extended calculations for both positive and negative parity 
levels up to 13 MeV in excitation. An excellent test for these extended 
calculations was their comparison with the three particle transfer reactions 
$^{12}$C($^{7}$Li,$\alpha$) \cite{Tse73} and $^{12}$C($^{6}$Li,$^{3}$He) 
\cite{Bin75} where both reactions selectively populate states including an 
especially strongly populated one at 10.69 MeV with much more strength than 
would be consistent with the known $\unitfrac{3}{2}^{-}$ level. The large 
angular momentum mismatch of the ($^{6}$Li,$^{3}$He) reaction favored the 
population of a high spin state, which was then matched to a 
$\unitfrac{9}{2}^{+}$ state calculated by Lie and Engeland \cite{Lie76} close 
to this excitation energy and  having a large 3p-4h component. Concurrently, 
a $^{14}$C(p,$\gamma$) study discovered a new resonance at 10.693 MeV in 
excitation and gave it a $\unitfrac{9}{2}^{+}$ assignment, and its decay 
reported in Ref. \cite{Beu75} was subsequently used to demonstrate that three 
particle transfer reactions \cite{Har79} do indeed strongly populate this 
3p-4h, $\unitfrac{9}{2}^{+}$ state as predicted by theory.

This work reports new data for the $^{14}$N(d,p) reaction taken at a 
bombarding energy of 16 MeV to locate all narrow neutron single particle 
states up to 15 MeV in excitation. It continues the ideas of the early 
shell model theoretical studies with the goal of testing a modern calculation 
against the known level structure of  $^{15}$N and  then  to use this 
calculation along with the new $^{14}$N(d,p) data to identify all single 
particle levels up to 12 MeV in excitation. These firm spin parity assignments 
are then used to determine their mirror levels in the lesser known nucleus 
$^{15}$O \cite{Ajz91}. The analysis of the present single neutron  data and 
that from a recently published $^{14}$N($^{3}$He,d) proton transfer study 
\cite{Ber02} are combined to determine the 2s$_{\unitfrac{1}{2}}$ and 
1d$_{\unitfrac{5}{2}}$  centroid energies in the $^{15}$N$-^{15}$O mirror pair. 
With these results, a reasonable estimate is made for the concentration of 
the 1d$_{\unitfrac{3}{2}}$ strength in these nuclei. The single particle 
centroid energies of the mass 15 nuclei are particularly interesting because 
they are in the cross over region where the 2s$_{\unitfrac{1}{2}}$ orbit lies 
below the 1d$_{\unitfrac{5}{2}}$ orbit in $^{13}$C, and above it in $^{17}$O. 
An analysis of an early extensive $^{14}$N(d,p) study \cite{Phi69} showed 
these orbits to be almost degenerate in $^{15}$N. However, the published 
spectrum showed possible single particle states for which no analysis of 
spectroscopic strength was carried out, suggesting the possibility that some 
strength for these orbits might lie at a higher excitation energy than 
studied to date. The present higher energy (d,p)  work  was designed to 
search for this possible missing strength.  The extraction of these centroid 
energies adds to our knowledge of the evolution of the s-d shell orbits as 
a function of proton and neutron number, the importance of which has been 
detailed in a recent publication of Hoffman {\em et al.} \cite{Hof14}.

\begin{figure*}[htb]
\centering
\includegraphics[width=1.5\columnwidth]{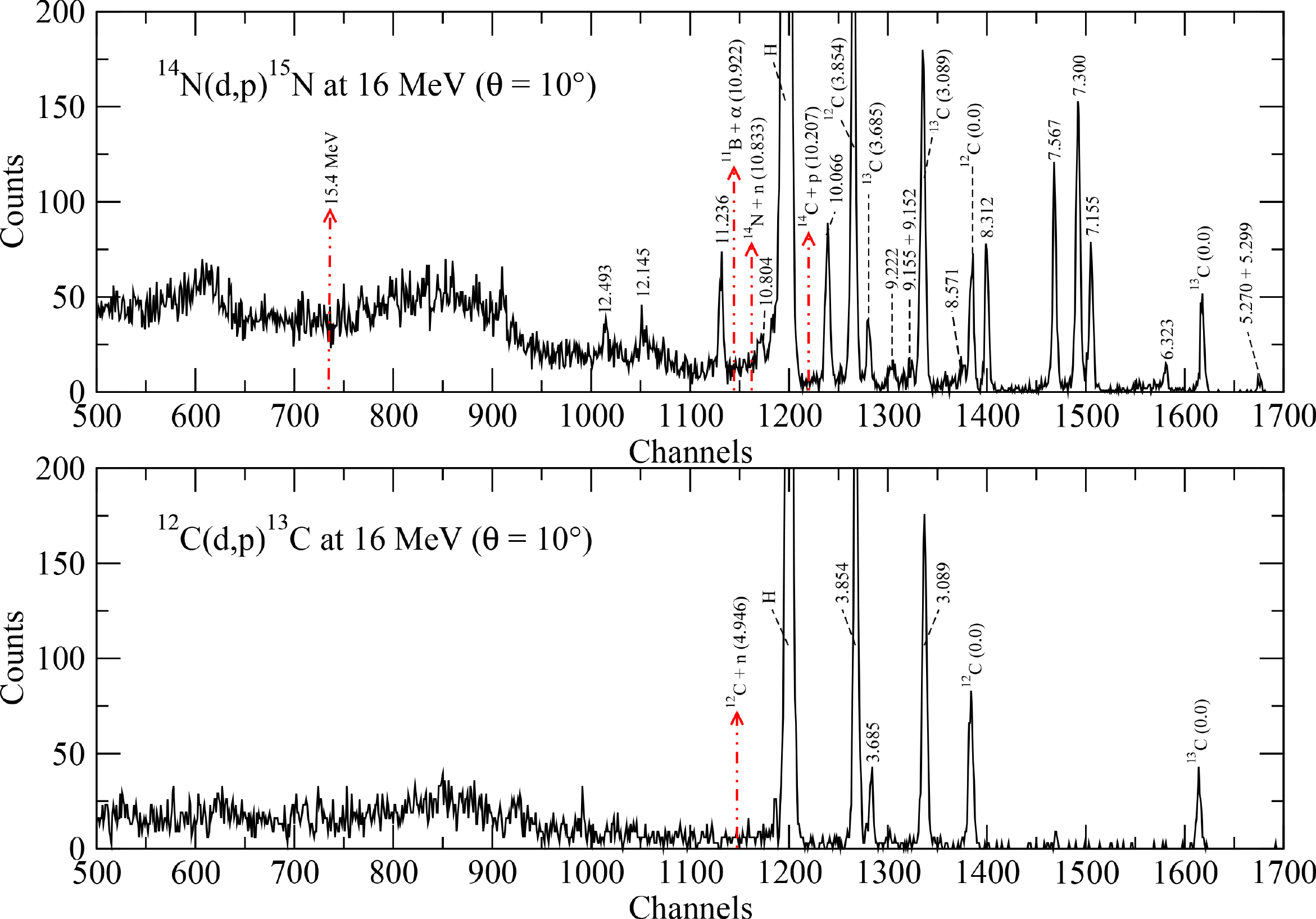}
\caption{(color online) Typical spectra from the present work, shows the 
$^{14}$N(d,p)$^{15}$N and $^{12}$C(d,p)$^{13}$C spectra at 10$^{\circ}$ taken
with a beam energy of 16 MeV. The $^{12}$C(d,p) spectra was gainshifted to match
the peak height of the 3.854 MeV state in $^{13}$C to show matching peaks. 
Separation energies have also been added to each spectra.}
\label{FIGURE:15N-spectra}
\end{figure*}

\section{\label{sec:level1}Experimental Procedure}
Cross sections for the $^{14}$N(d,p)$^{15}$N reaction were obtained by 
bombarding a melamine (C$_{3}$H$_{6}$N$_{6}$) target of about 
300~$\unitfrac{\mu \text{g}}{\text{cm}^{2}}$ on a 
20~$\unitfrac{\mu \text{g}}{\text{cm}^{2}}$ carbon backing with a 16 MeV 
deuteron beam produced by the FSU tandem-linac accelerator combination. 
Data were also collected on a carbon target at the same angles as for the 
melamine to separate its contribution to the higher excitation $^{15}$N 
spectra from the melamine, the main experimental thrust of the current work. 
A $\Delta$E-E silicon detector telescope was used to measure the outgoing 
protons and deuterons and a single silicon detector on the opposite side of 
the incoming beam served as a monitor of the target conditions. The deuteron 
beam current was limited to 15~nA to make certain that there was no loss of 
target due to beam heating during the data collection. Elastic scattering and 
transfer data were also taken at deuteron beam energies of 9, 10, and 12 MeV 
to extract absolute cross sections and to check these cross sections with 
previously published (d,p) data \cite{Phi69}. To make use of previously 
published $^{12}$C and $^{14}$N deuteron elastic scattering data to establish 
the absolute cross section for the (d,p) reaction, it was necessary to adopt 
a procedure to separate the two peak yields since the elastic peak contained 
yield from both nuclei, most notably at forward angles. To extract the carbon 
yield from the total peak, the ratio of the yield for the 4.4 MeV first 
excited state in $^{12}$C to that for the $^{12}$C ground state for the 
deuteron scattering by the carbon target was found at all measured angles. The 
elastic nitrogen yield was determined by extracting the 4.4 MeV yield from the 
melamine target and then using the yield ratio from the carbon target to 
subtract its contribution to the total nitrogen plus carbon elastic yield. 
For the angles 25$^{\circ}$, 30$^{\circ}$ and 35$^{\circ}$ it was possible to 
separate the yields from carbon and nitrogen which provided a direct check 
on the ratio technique. The relative elastic scattering angular distributions 
for both carbon and nitrogen were normalized to optical model calculations 
based on previously published elastic scattering data \cite{BusJar74} for 
both nuclei to extract a normalization constant that establishes the absolute 
cross section for the (d,p) transfer reactions The error in the absolute 
cross sections in Ref. \cite{BusJar74} is 15\% which is taken as the 
absolute error here. 

A typical $^{14}$N(d,p) spectrum showing the population of $^{15}$N states up 
to 15 MeV in excitation is displayed in Figure \ref{FIGURE:15N-spectra}. Also 
shown are the energies for the various particle decay thresholds for $^{15}$N. 
Note also in the spectrum the contribution from the $^{12}$C(d,p) peaks and 
the strong peak arising from the hydrogen in the target. Figure 
\ref{FIGURE:15N-spectra} also shows the $^{12}$C(d,p) spectrum below that 
from the melamine demonstrating that the narrow peaks above 10 MeV are indeed 
from the $^{14}$N(d,p) reaction. All the narrow peaks below 11.5 MeV in 
excitation reported earlier in Phillips and Jacobs \cite{Phi69} were observed 
in the present work including the narrow peaks at 10.066 and 11.236 MeV, 
whereas those at 12.145 and 12.493 MeV were not. The 11.236 MeV angular 
distribution was not included in the analysis of Ref. \cite{Phi69}. It was 
not possible to extract an angular distribution for the small peak in the 
spectrum at 10.80 MeV in excitation because of its weak population and its 
being obscured at many angles by the nearby peak from the hydrogen in the 
target.  In contrast with various multiparticle transfer reactions that 
selectively populate narrow peaks up to 20 MeV in excitation in $^{15}$N, 
no strong isolated peaks are observed in the $^{14}$N(d,p) spectrum above 
12.5 MeV in excitation. While there appears to be a broad peak around 14.5 MeV 
in excitation, it was not possible to prove conclusively that it was not 
from the known broad $\unitfrac{3}{2}^{+}$ state at 8.2 MeV in $^{13}$C 
\cite{Ohn85}. Data were taken in the laboratory angular range from 
10$^{\circ}$ to 35$^{\circ}$ in 5$^{\circ}$ steps to produce angular 
distributions for extracting the orbital angular momentum transfer and the 
single particle spectroscopic factors that provide a measure of the neutron 
single particle strength for a given state. The size of the relative errors 
in the angular distribution data are either smaller or equal to the size 
of the data points or are shown.

\section{\label{sec:level1}Shell Model Calculations}
\begin{figure}[htb]
\centering
\includegraphics[width=\columnwidth]{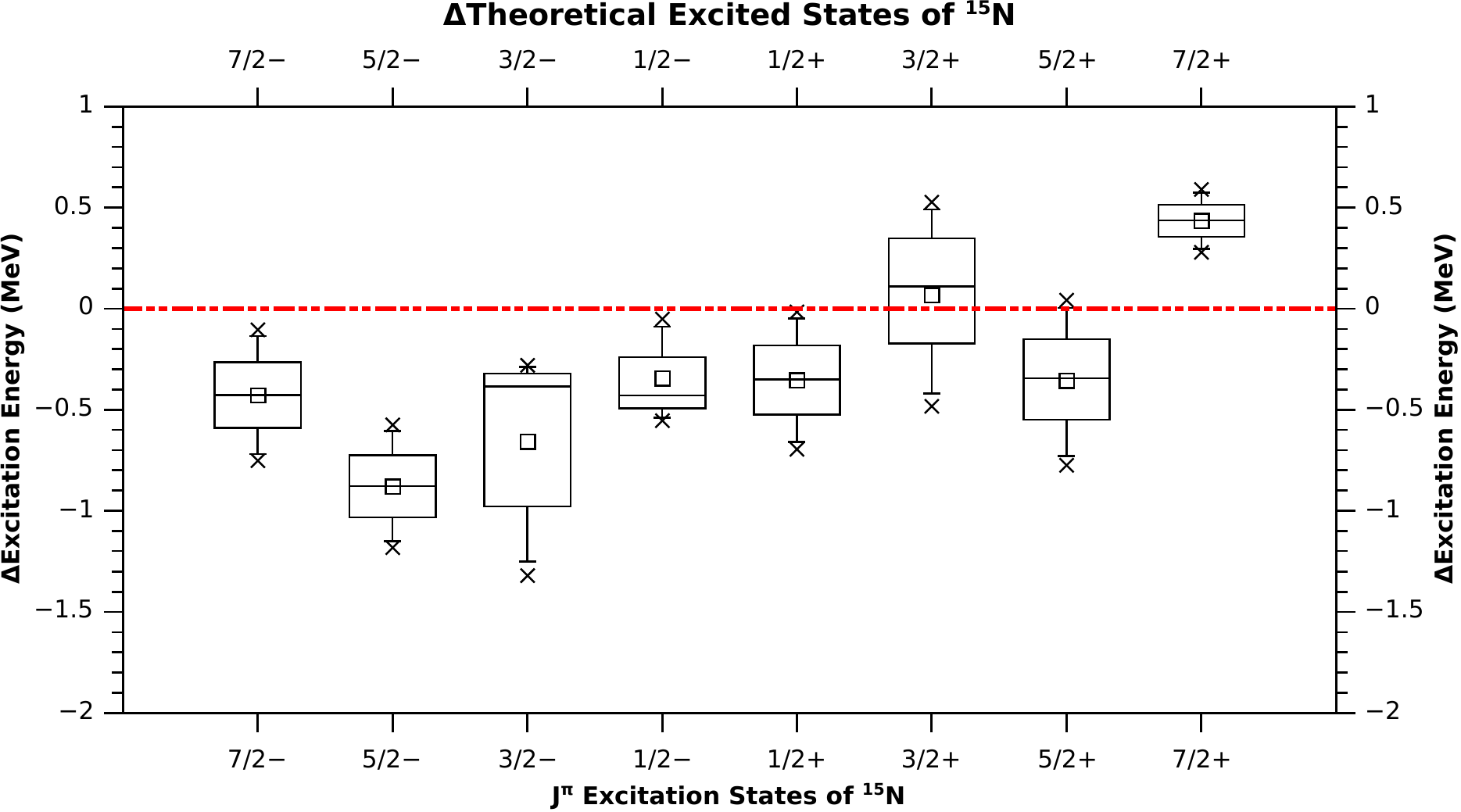}
\caption{(color online) Distribution of the theoretical calculations in 
relation to the experimentally observed excitation energies in $^{15}$N. 
The dashed line at zero denotes a match in energy values.}
\label{FIGURE:TheoryDist}
\end{figure}

The present work reports the results of new $^{15}$N shell model calculations 
that used an unrestricted 1p-2s1d shell valence space with an interaction 
Hamiltonian taken from the work of Utsuno and Chiba \cite{Uts01}. Time 
dependent and traditional shell model procedures were used with the computer 
code {\sc Cosmo} \cite{Vol09} to perform the calculations. The known and 
calculated states in $^{15}$N are given in the first two columns of Table 
\ref{TABLE:15N-15O}, where it can be seen that all known levels can be paired 
with calculated ones. The number in parenthesis besides that for a given 
spin and parity refers to the theoretical level ordering so that for 
example $\unitfrac{5}{2}^{+}\;(3)$ is the third $\unitfrac{5}{2}^{+}$ level 
and its known partner is found at 9.155 MeV. Figure \ref{FIGURE:TheoryDist} 
displays the difference between the known and calculated levels. The energies 
of the calculated positive parity states are fairly well predicted by the 
calculations, but the energies of the negative parity states are higher than 
their corresponding experimental ones.

\begin{table*}[htb]
\centering
\caption{Assigned Mirror States in $^{15}$N and $^{15}$O with Energy Differences}
  \begin{tabular}{@{}l*{8}{c}r@{}}
    \hline\hline
    \multicolumn{2}{c}{$^{15}$N} & \; &\multicolumn{2}{c}{Theoretical 
$^{15}$N$^{\dagger}$} &\; & \multicolumn{2}{c}{$^{15}$O}  & $\Delta$E (MeV) 
& $\Delta$E (MeV)\Tstrut\\
    \cline{1-2} \cline {4-5} \cline{7-8}
    E (MeV)    & $J^{\pi}$ & & E (MeV) & $J^{\pi}$\hspace{.5em}$_{(n)}$ & & 
E (MeV) & $J^{\pi}$ & $^{15}$N & $^{15}$N$-^{15}$O\phantom{a}\Tstrut\Bstrut \\
    \hline
    5.270  & $\unitfrac{5}{2}^{+}$ & & 5.227 & 
$\unitfrac{5}{2}^{+}$\hspace{.05em}$_{(1)}$ && 5.241 & $\unitfrac{5}{2}^{+}$ 
& 0.043 & 0.029\Tstrut \\
    5.299  & $\unitfrac{1}{2}^{+}$ & & 5.994 & 
$\unitfrac{1}{2}^{+}$\hspace{.05em}$_{(1)}$ && 5.183 & $\unitfrac{1}{2}^{+}$ 
& $-0.695$\phantom{$-$} & 0.166 \\
    6.324  & $\unitfrac{3}{2}^{-}$ & & 6.602 & 
$\unitfrac{3}{2}^{-}$\hspace{.05em}$_{(1)}$ && 6.176 & $\unitfrac{3}{2}^{-}$ 
& $-0.279$\phantom{$-$} & 0.148 \\
    7.155  & $\unitfrac{5}{2}^{+}$ & & 7.370 & 
$\unitfrac{5}{2}^{+}$\hspace{.05em}$_{(2)}$ && 6.859 & $\unitfrac{5}{2}^{+}$ 
& $-0.215$\phantom{$-$} & 0.296 \\
    7.300  & $\unitfrac{3}{2}^{+}$ & & 6.772 
& $\unitfrac{3}{2}^{+}$\hspace{.05em}$_{(1)}$ && 6.793 & $\unitfrac{3}{2}^{+}$ 
& 0.528 & 0.508 \\
    7.567  & $\unitfrac{7}{2}^{+}$ & & 6.976 & 
$\unitfrac{7}{2}^{+}$\hspace{.05em}$_{(1)}$ && 7.276 & $\unitfrac{7}{2}^{+}$ 
& 0.591 & 0.291 \\
    8.312  & $\unitfrac{1}{2}^{+}$ & & 8.327 & 
$\unitfrac{1}{2}^{+}$\hspace{.05em}$_{(2)}$ && 7.557 & $\unitfrac{1}{2}^{+}$ 
& $-0.015$\phantom{$-$} & 0.756 \\
    8.571  & $\unitfrac{3}{2}^{+}$ & & 8.638 & 
$\unitfrac{3}{2}^{+}$\hspace{.05em}$_{(2)}$ && 8.284 & $\unitfrac{3}{2}^{+}$ 
& $-0.067$\phantom{$-$} & 0.287 \\
    9.050  & $\unitfrac{1}{2}^{+}$ & & 9.400 & 
$\unitfrac{1}{2}^{+}$\hspace{.05em}$_{(3)}$ && 8.743 & $\unitfrac{1}{2}^{+}$ 
& $-0.350$\phantom{$-$} & 0.307 \\
    9.152  & $\unitfrac{3}{2}^{-}$ & & 9.474 & 
$\unitfrac{3}{2}^{-}$\hspace{.05em}$_{(2)}$ && 8.922 & $\unitfrac{3}{2}^{-}$ & 
$-0.322$\phantom{$-$} & 0.230 \\
    9.155 & $\unitfrac{5}{2}^{+}$ & & 9.929 & 
$\unitfrac{5}{2}^{+}$\hspace{.05em}$_{(3)}$ && 8.922 & $\unitfrac{5}{2}^{+}$ & 
$-0.774$\phantom{$-$} & 0.233 \\
    9.222  & $\unitfrac{1}{2}^{-}$ & & 9.273 & 
$\unitfrac{1}{2}^{-}$\hspace{.05em}$_{(2)}$ && 8.982 & $\unitfrac{1}{2}^{-}$ & 
$-0.051$\phantom{$-$} & 0.242 \\
    9.760  & $\unitfrac{5}{2}^{-}$ & & 10.335 & 
$\unitfrac{5}{2}^{-}$\hspace{.05em}$_{(1)}$&& 9.488 & $\unitfrac{5}{2}^{-}$ & 
$-0.575$\phantom{$-$} & 0.272 \\
    9.829  & $\unitfrac{7}{2}^{-}$ & & 10.580 & 
$\unitfrac{7}{2}^{-}$\hspace{.05em}$_{(1)}$&& 9.660 & $\unitfrac{7}{2}^{-}$ & 
$-0.751$\phantom{$-$} & 0.169 \\
    9.925  & $\unitfrac{3}{2}^{-}$ & & 10.310 & 
$\unitfrac{3}{2}^{-}$\hspace{.05em}$_{(3)}$&& 9.609 & $\unitfrac{3}{2}^{-}$ & 
$-0.385$\phantom{$-$} & 0.316 \\
    10.066 & $\unitfrac{3}{2}^{+}$ & & 9.779 & 
$\unitfrac{3}{2}^{+}$\hspace{.05em}$_{(3)}$ && 9.484  & $\unitfrac{3}{2}^{+}$ & 
0.287 & 0.582 \\
    10.450 & $\unitfrac{5}{2}^{-}$ & & 11.631 & 
$\unitfrac{5}{2}^{-}$\hspace{.05em}$_{(2)}$&& 10.290 & $\unitfrac{5}{2}^{-}$ & 
$-1.181$\phantom{$-$}& 0.160 \\
    10.533 & $\unitfrac{5}{2}^{+}$ & & 11.005 & 
$\unitfrac{5}{2}^{+}$\hspace{.05em}$_{(4)}$&& 10.300 & $\unitfrac{5}{2}^{+}$ & 
$-0.472$\phantom{$-$} & 0.233 \\
    10.693 & $\unitfrac{9}{2}^{+}$ & & 12.292 & 
$\unitfrac{9}{2}^{+}$\hspace{.05em}$_{(1)}$&& 10.461 & $\unitfrac{9}{2}^{+}$ & 
$-1.600$\phantom{$-$}& 0.229 \\
    10.702 & $\unitfrac{3}{2}^{-}$ & & 12.022 & 
$\unitfrac{3}{2}^{-}$\hspace{.05em}$_{(4)}$&& 10.480 & $\unitfrac{3}{2}^{-}$ & 
$-1.320$\phantom{$-$}& 0.222 \\
    10.804 & $\unitfrac{3}{2}^{+}$ & & 11.286 & 
$\unitfrac{3}{2}^{+}$\hspace{.05em}$_{(4)}$&& 10.506 & $\unitfrac{3}{2}^{+}$ & 
$-0.482$\phantom{$-$}& 0.298 \\
    11.236 & $\unitfrac{7}{2}^{+}$ & & 10.956 & 
$\unitfrac{7}{2}^{+}$\hspace{.05em}$_{(2)}$ && 10.917 & $\unitfrac{7}{2}^{+}$ & 
0.280 & 0.318 \\
    11.293 & $\unitfrac{1}{2}^{-}$ & & 11.846 & 
$\unitfrac{1}{2}^{-}$\hspace{.05em}$_{(3)}$&& 11.025 & $\unitfrac{1}{2}^{-}$ & 
$-0.553$\phantom{$-$} & 0.267\Tstrut\Bstrut \\
    \hline
    Avg Diff & & & & &&&& $-0.365$\phantom{$-$}& 0.285\\
    \hline\hline
    \multicolumn{10}{@{}l}{\scriptsize $\dagger$\hspace{.4em} $(n)$ denotes 
the order which the $J^{\pi}$ state appeared in the calculations.}
\end{tabular}
\label{TABLE:15N-15O}
\end{table*}

\section{\label{sec:level1}DWBA Calculations}
Zero-range distorted-wave Born approximation (ZR-DWBA) calculations were 
performed using the code {\sc Dwuck4} \cite{KunDWBA}. The {\sc Dwuck4} code 
enables the use of the technique of Vincent and Fortune \cite{Vin73} for 
handling the target-like overlap when the final state in the residual 
nucleus is unbound and the incorporation of a non-locality correction factor 
$\beta$ for the distorted waves and transferred particle bound state. 
Deuteron and proton optical model potentials were taken from the study of 
Phillips and Jacobs \cite{Phi69}. The 
\bracket{^{15}\text{N}}{^{14}\text{N + n}} overlaps were calculated using 
binding potentials of Woods-Saxon form and parameters r$_{0} = 1.25$ fm, 
a$_{0} = 0.65$ fm and a Thomas spin-orbit term with strength $\lambda = 25$. 
A value of D$_{0}^{2} = 1.55\times10^{4}\ \text{MeV}^{2}\ \text{fm}^{3}$ and 
a finite-range correction factor of 0.621 fm were used, as recommended in 
the {\sc Dwuck4} manual for (d,p) reactions. Non-locality corrections 
\cite{Per62} with $\beta = 0.54$ fm for the deuteron and $\beta = 0.85$ fm 
for the proton and transferred neutron were also applied.

For unbound states, calculations assuming the transferred neutron to be in 
an $\ell = 0$ state employed the weak binding energy approximation (WBEA), 
i.e. the \bracket{^{15}\text{N}}{^{14}\text{N + n}} overlap was calculated 
assuming a binding energy of 0.1 MeV, although the correct excitation energy 
of the state was used for the ``kinematic'' part of the calculation. A similar 
procedure also had to be adopted for the 12.145 MeV and 12.493 MeV unbound 
states when the transferred neutron was assumed to be in an $\ell = 1$ state 
since it was not possible to find resonances for these states under this 
assumption, the single-particle widths being too wide. However, the 
Vincent-Fortune technique \cite{Vin73} was employed to calculate the 
$\ell = 2$ angular distributions for these two states and both the $\ell = 1$ 
and $\ell = 2$ angular distributions for all the other unbound states.  
For those calculations where the WBEA had to be used no non-locality 
corrections were employed since their influence on the result is likely to 
be smaller than the effect of using \bracket{^{15}\text{N}}{^{14}\text{N + n}} 
overlaps calculated under the somewhat crude assumptions of the WBEA. 
Normalization factors were extracted for each transition by multiplying each 
calculated transfer cross section by a number until the two magnitudes 
matched. This normalization factor is then the single particle spectroscopic 
factor $C^{2}S$. Transitions to $\unitfrac{3}{2}^{+}$ states can proceed by 
either $\ell = 0$ or $\ell = 2$ transfers and in these cases the 
normalizations for the two transfers were varied until the value of 
$\chi^{2}$ that combined the two calculations was minimized. 

\section{\label{sec:level1}Results}

\begin{figure}[H]
\centering
\includegraphics[width=.8\columnwidth]{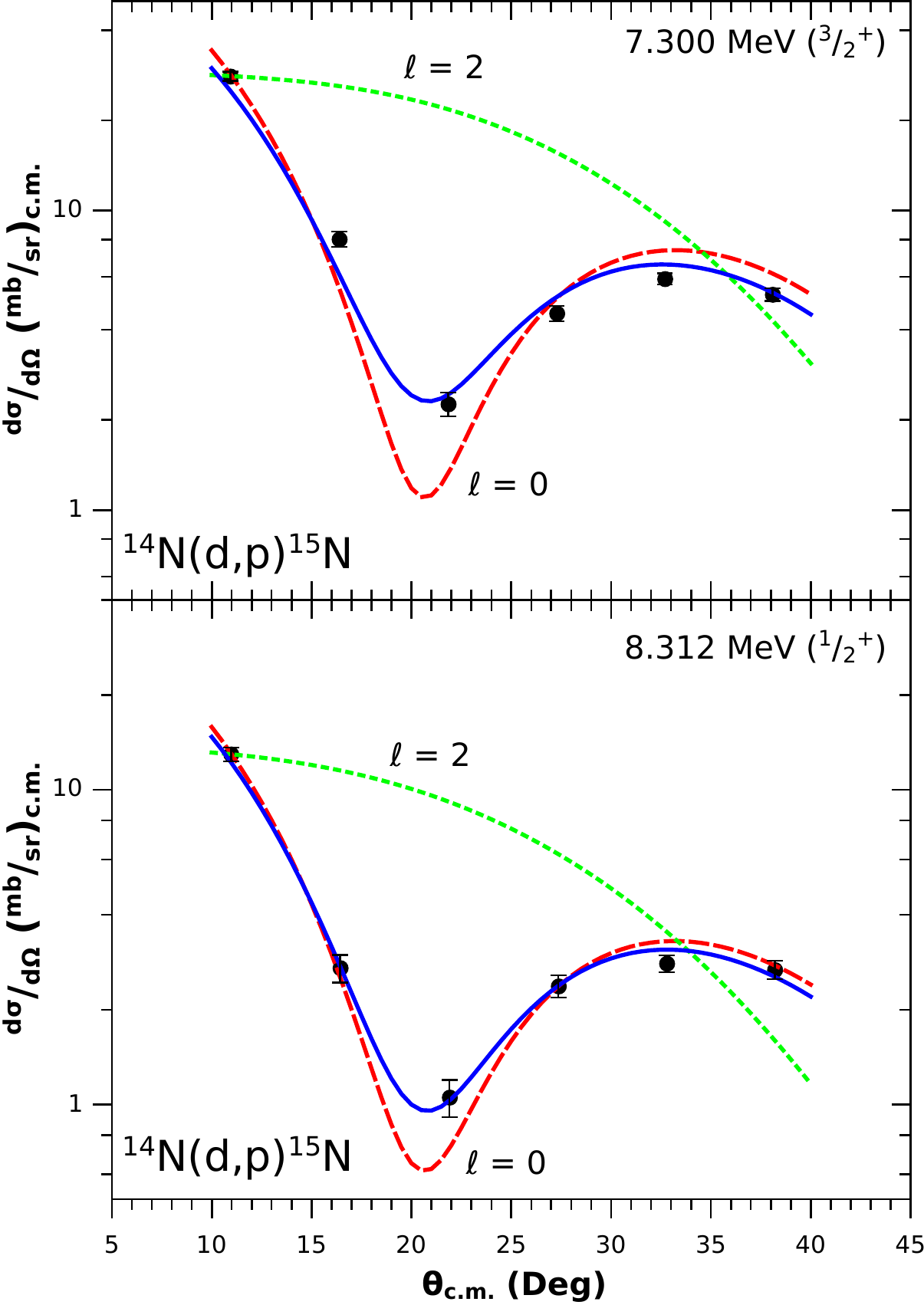}
\caption{(color online) Angular distribution for states in $^{15}$N which 
were determined to be $\ell = 0$ dominant.}
\label{FIGURE:L0_Dom}
\end{figure}

\begin{figure}[htb]
\centering
\includegraphics[width=.8\columnwidth]{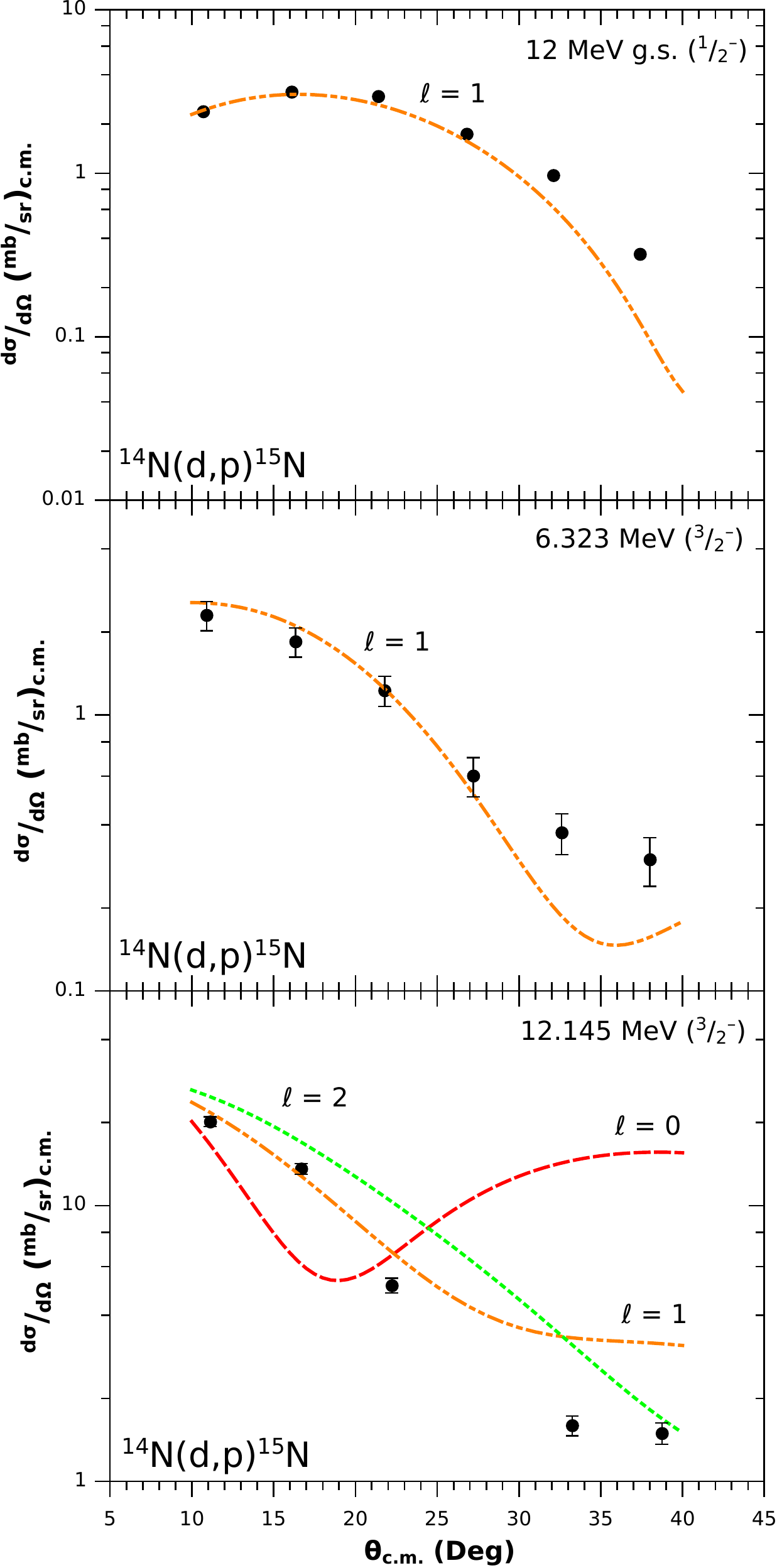}
\caption{(color online) Angular distribution for states in $^{15}$N which 
were determined to be $\ell = 1$ dominant.}
\label{FIGURE:L1_Dom}
\end{figure}

\begin{figure*}[htb]
\centering
\includegraphics[width=1.5\columnwidth]{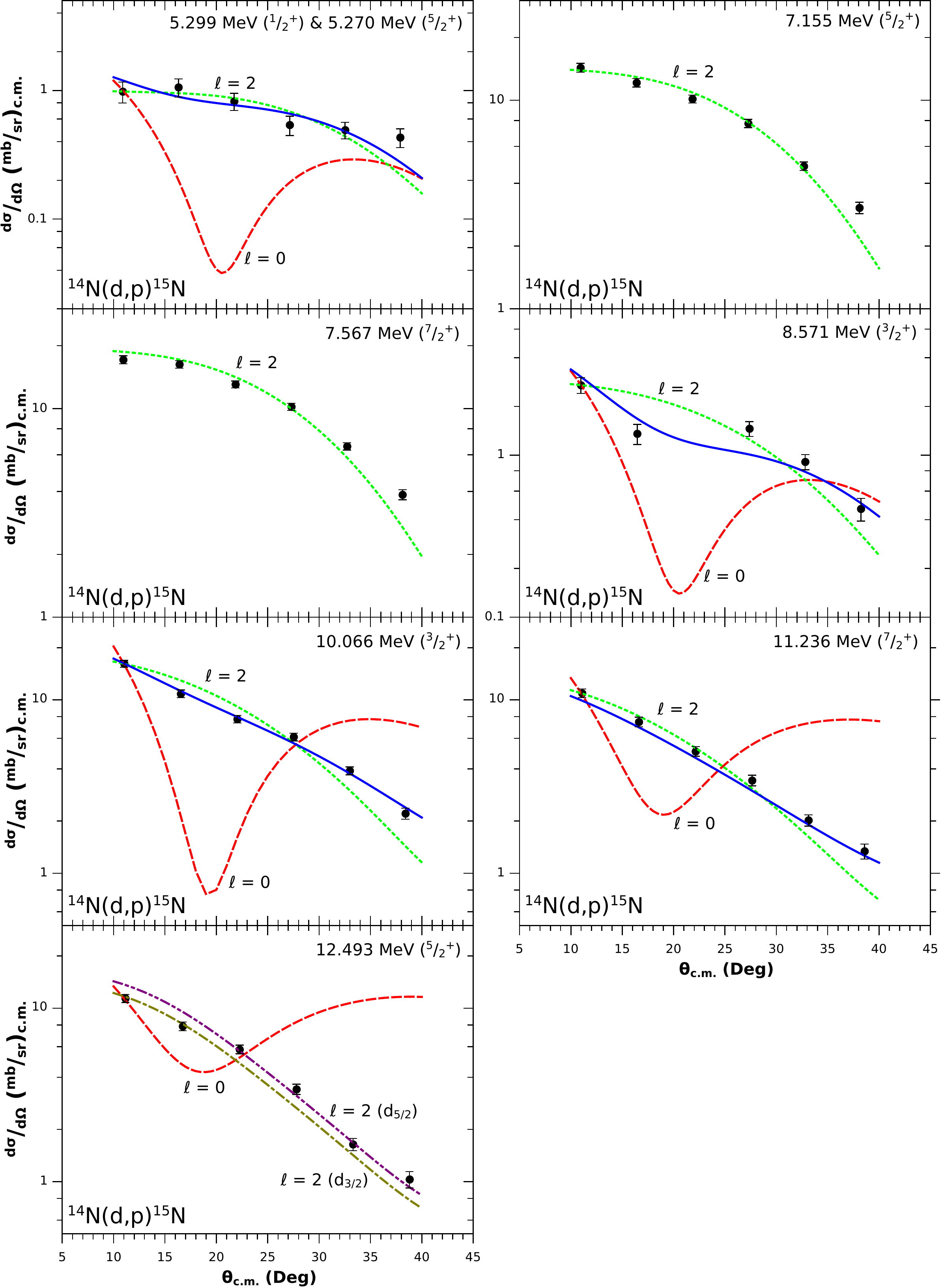}
\caption{(color online) Angular distribution for states in $^{15}$N which 
were determined to be $\ell = 2$ dominant.}
\label{FIGURE:L2_Dom}
\end{figure*}

\begin{table}[htb]
\centering
\caption{Spectroscopic Factors for the $^{14}$N(d,p)$^{15}$N Reaction}
  \begin{tabular}{@{}l c c c c c r@{}}
    \hline\hline
    \ & & & & \multicolumn{3}{c}{$C^{2}S$}\Tstrut\\
    \cline{5-7}
    E (MeV) & $\ell (\hbar)$ & $J^{\pi}$ & Orbit & 16 MeV & Theoretical & 
9 MeV$^{a}$\Tstrut\Bstrut \\ \hline
    0.00 (g.s.) & 1 & $\unitfrac{1}{2}^{-}$ & p$_{\unitfrac{1}{2}}$ & 1.31 & 
1.14 & ---\phantom{aaa}\\
    5.280  & 0 & $\unitfrac{1}{2}^{+}$ & s$_{\unitfrac{1}{2}}$ & 0.03(02) & 
0.39 & $<0.05$\phantom{(00)}\Tstrut \\
    \      & 2 & $\unitfrac{5}{2}^{+}$ & d$_{\unitfrac{5}{2}}$ & 0.14(01) & 
0.11 & $<0.05$\phantom{(00)}\\
    6.323  & 1 & $\unitfrac{3}{2}^{-}$ & p$_{\unitfrac{3}{2}}$ & 0.22(01) & 
--- & 0.10(02)\\
    7.155  & 2 & $\unitfrac{5}{2}^{+}$ & d$_{\unitfrac{5}{2}}$ & 1.06(02) & 
0.65 & 0.88(03)\\
    7.300  & 0 & $\unitfrac{3}{2}^{+}$ & s$_{\unitfrac{1}{2}}$ & 0.98(03) & 
0.72 & 0.89(04)\\
    \      & 2 &                    \ & d$_{\unitfrac{5}{2}}$ & 0.19(03) & 
--- & 0.07(05)\\
    7.567  & 2 & $\unitfrac{7}{2}^{+}$ & d$_{\unitfrac{5}{2}}$ & 0.96(02) & 
0.73 & 0.87(01)\\
    8.312  & 0 & $\unitfrac{1}{2}^{+}$ & s$_{\unitfrac{1}{2}}$ & 1.10(05) & 
0.64 & 1.02(04)\\
    \      & 2 &                    \ & d$_{\unitfrac{3}{2}}$ & 0.10(04) & 
--- & $<0.09$\phantom{(00)}\\
    8.571  & 0 & $\unitfrac{3}{2}^{+}$ & s$_{\unitfrac{1}{2}}$ & 0.07(02) & 
--- & 0.02(01)\\
    \      & 2 &                    \ & d$_{\unitfrac{5}{2}}$ & 0.13(02) & 
0.20 & 0.12(03)\\
    10.066 & 0 & $\unitfrac{3}{2}^{+}$ & s$_{\unitfrac{1}{2}}$ & 0.32(04) & 
--- & 0.32(08)\\
    \      & 2 &                      & d$_{\unitfrac{5}{2}}$ & 0.65(02) & 
0.55 & 0.48(08)\\
    11.236 & 2 & $\unitfrac{7}{2}^{+}$ & d$_{\unitfrac{5}{2}}$ & 0.20(01) & 
--- & ---\phantom{aaa}\\
    12.493 & 2 & $\unitfrac{5}{2}^{+}$ & d$_{\unitfrac{3}{2}}$ & 0.28(01) & 
--- & ---\phantom{aaa}\\
    \      & 2 &                    \ & d$_{\unitfrac{5}{2}}$ & 0.30(01) & 
--- & ---\phantom{aaa}\Bstrut\\
    \hline\hline
    \multicolumn{6}{@{}l}{\scriptsize $a$\hspace{.52em} Data extracted 
from \cite{Phi69}.} \\
  \end{tabular}
\label{TABLE:SF-15N}
\end{table}
\ \\
The current experimental set up was optimized to look for structure in the 
experimental spectrum above 10 MeV in excitation so that no data were taken 
for the ground state transition. For completeness in the current analysis, 
data taken by Schiffer {\em et al.} \cite{Sch67} in a survey of ground state 
(d,p) transitions in 1p shell nuclei at a deuteron bombarding energy of 
12 MeV  were also analyzed. As can be seen in Figs. 
\ref{FIGURE:L0_Dom}$-$\ref{FIGURE:L2_Dom}, the angular distributions are 
well described by the DWBA calculations. The descriptions of both the 
7.300 $\left(\unitfrac{3}{2}^{+}\right)$ and 8.312 
$\left(\unitfrac{1}{2}^{+}\right)$ angular distributions are improved by 
slight addition of an $\ell = 2$ transfer to the dominant $\ell = 0$ transfer. 
An $\ell = 2$ contribution to the 8.312 MeV transition can only occur from 
a 1d$_{\unitfrac{3}{2}}$ neutron configuration which would yield a component 
of this orbit much lower than expected but its addition at less than 10\% 
of the dominant $\ell = 0$ component is determined primarily by the small 
cross section to this state at the angular distribution minimum which results 
in a large error associated with the extraction of this component.

\begin{table}[!hb]
\centering
\caption{Currently reported states in $^{15}$N and $^{15}$O$^{a}$}
   \begin{tabular}{@{}l c c c r@{}}
     \hline\hline
     \multicolumn{2}{c}{$^{15}$N} & \; & \multicolumn{2}{c}{$^{15}$O}\Tstrut\\
     \cline{1-2} \cline{4-5}
     E (MeV) & $J^{\pi}$ & $\quad$ & E (MeV) & $J^{\pi}$\Tstrut\Bstrut \\
     \hline
     5.270 & $\unitfrac{5}{2}^{+}$ & & 5.241 & $\unitfrac{5}{2}^{+}$ \\ 
     5.299 & $\unitfrac{1}{2}^{+}$ & & 5.183 & $\unitfrac{1}{2}^{+}$ \\
     6.324 & $\unitfrac{3}{2}^{-}$ & & 6.176 & $\unitfrac{3}{2}^{-}$ \\
     7.155 & $\unitfrac{5}{2}^{+}$ & & 6.859 & $\unitfrac{5}{2}^{+}$ \\
     7.300 & $\unitfrac{3}{2}^{+}$ & & 6.793 & $\unitfrac{3}{2}^{+}$ \\
     7.567 & $\unitfrac{7}{2}^{+}$ & & 7.276 & $\unitfrac{7}{2}^{+}$ \\
     8.312 & $\unitfrac{1}{2}^{+}$ & & 7.557 & $\unitfrac{1}{2}^{+}$ \\
     8.571 & $\unitfrac{3}{2}^{+}$ & & 8.284 & $\unitfrac{3}{2}^{+}$ \\
     9.050 & $\unitfrac{1}{2}^{+}$ & & 8.743 & $\unitfrac{1}{2}^{+}$ \\
     9.152 & $\unitfrac{3}{2}^{-}$ & & ---   & ---\phantom{a}        \\
     9.155 & $\unitfrac{5}{2}^{+}$ & & 8.922 & $\unitfrac{5}{2}^{+}$ \\
     \phantom{a}--- & ---          && 8.922 & $\unitfrac{1}{2}^{+}$  \\
     9.222 & $\unitfrac{1}{2}^{-}$ & & 8.982 & 
$\left(\unitfrac{1}{2}^{-}\right)$ \\
     \phantom{a}--- & ---         & & 9.484 & 
$\left(\unitfrac{3}{2}^{+}\right)$ \\
     9.760 & $\unitfrac{5}{2}^{-}$ & & 9.488 & $\unitfrac{5}{2}^{-}$ \\
     9.829 & $\unitfrac{7}{2}^{-}$ & & 9.660 & %
                     $\left(\unitfrac{7}{2}^{-},\ \unitfrac{9}{2}^{-}\right)$ \\
     9.925 & $\unitfrac{3}{2}^{-}$ & & 9.609 & $\unitfrac{3}{2}^{-}$ \\
     10.066 & $\unitfrac{3}{2}^{+}$ & & --- & ---\phantom{a} \\
     10.450 & $\unitfrac{5}{2}^{-}$ && 10.290 & 
$\left(\unitfrac{5}{2}^{-}\right)$ \\
     10.533 & $\unitfrac{5}{2}^{+}$ && 10.300 & $\unitfrac{5}{2}^{+}$\\
     10.693 & $\unitfrac{9}{2}^{+}$ && 10.461 & 
$\left(\unitfrac{9}{2}^{+}\right)$ \\
     10.702 & $\unitfrac{3}{2}^{-}$ && 10.480 & 
$\left(\unitfrac{3}{2}^{-}\right)$ \\
     10.804 & $\unitfrac{3}{2}^{+}$ && $\left( 10.506 \right)$ & 
$\left(\unitfrac{3}{2}^{+}\right)$ \\
     11.236 & \hspace{-1.75em}$\geq \unitfrac{3}{2}$ && 10.917 & 
$\unitfrac{7}{2}^{+}$\Bstrut \\
     \hline\hline
     \multicolumn{5}{@{}l}{\scriptsize $a$\hspace{.52em} Data extracted 
from \cite{Ajz91}.}
   \end{tabular}
   \label{TABLE:15N-O-Old}
\end{table} 

\ \\
The only calculated level below 11.5 MeV for which there is no known 
experimental equivalent is the $\unitfrac{7}{2}^{+}\;(2)$ predicted to lie at 
10.956 MeV. The angular distribution for the 11.236 peak is well described by 
an $\ell = 2$ transfer which is consistent with a spin of 
$\unitfrac{7}{2}^{+}$. This assignment would also agree with the most recent 
compilation for mass 15 \cite{Ajz91} which has the 11.236 MeV state listed as 
having a spin greater than or equal to $\unitfrac{3}{2}$.  Further support for 
this assignment can be found by considering the width of a possible single 
particle level in a Woods-Saxon potential well. If the spin of the 11.236 MeV 
level were either $\unitfrac{1}{2}^{+}$ or $\unitfrac{3}{2}^{+}$, both allowed 
for an $\ell = 2$ transfer, then they would have an $\ell = 0$ component which 
would require a width for this state of 300 keV or so, whereas the measured 
width is 3.3 keV.  If it were $\unitfrac{5}{2}^{+}$, then it would be the 
fifth $\unitfrac{5}{2}^{+}$ level and one sees that the $\unitfrac{5}{2}^{+}$ 
single particle strength is exhausted by this excitation energy, leaving 
$\unitfrac{7}{2}^{+}\;(2)$ as the remaining choice. The shell model 
calculation has a spectroscopic factor of 0.71 for $\unitfrac{7}{2}^{+}\;(1)$ 
and 0.135 for $\unitfrac{7}{2}^{+}\;(2)$ and 0.009 for 
$\unitfrac{7}{2}^{+}\;(3)$ again supporting the d$_{\unitfrac{5}{2}}$ 
$\unitfrac{7}{2}^{+}$ component for this level.

 The compilation \cite{Ajz91} shows a $\unitfrac{5}{2}^{+}$ state at 12.493 
MeV which  is consistent with the present $\ell = 2$ angular distribution 
analysis. However, whether it is a remnant of the 1d$_{\unitfrac{5}{2}}$ orbit 
or the beginning of the single particle strength for the 1d$_{\unitfrac{3}{2}}$ 
orbit cannot be determined from the present work. Comparison with the 
$^{12}$C(d,p) analysis of Ohnuma {\em et al.} \cite{Ohn85} suggests that it is 
the beginning of the 1d$_{\unitfrac{3}{2}}$ orbit. The present spectroscopic 
factors with the corresponding errors arising from the fit to the data along 
with those obtained by Phillips and Jacobs are given in Table 
\ref{TABLE:SF-15N}. A reanalysis of both the ground state and excited states 
from the previously published data in Refs. \cite{Phi69, Sch67} resulted in 
agreement between the two analyses giving considerable confidence in their 
values. 

\begin{figure}[htb]
\centering
\includegraphics[width=\columnwidth]{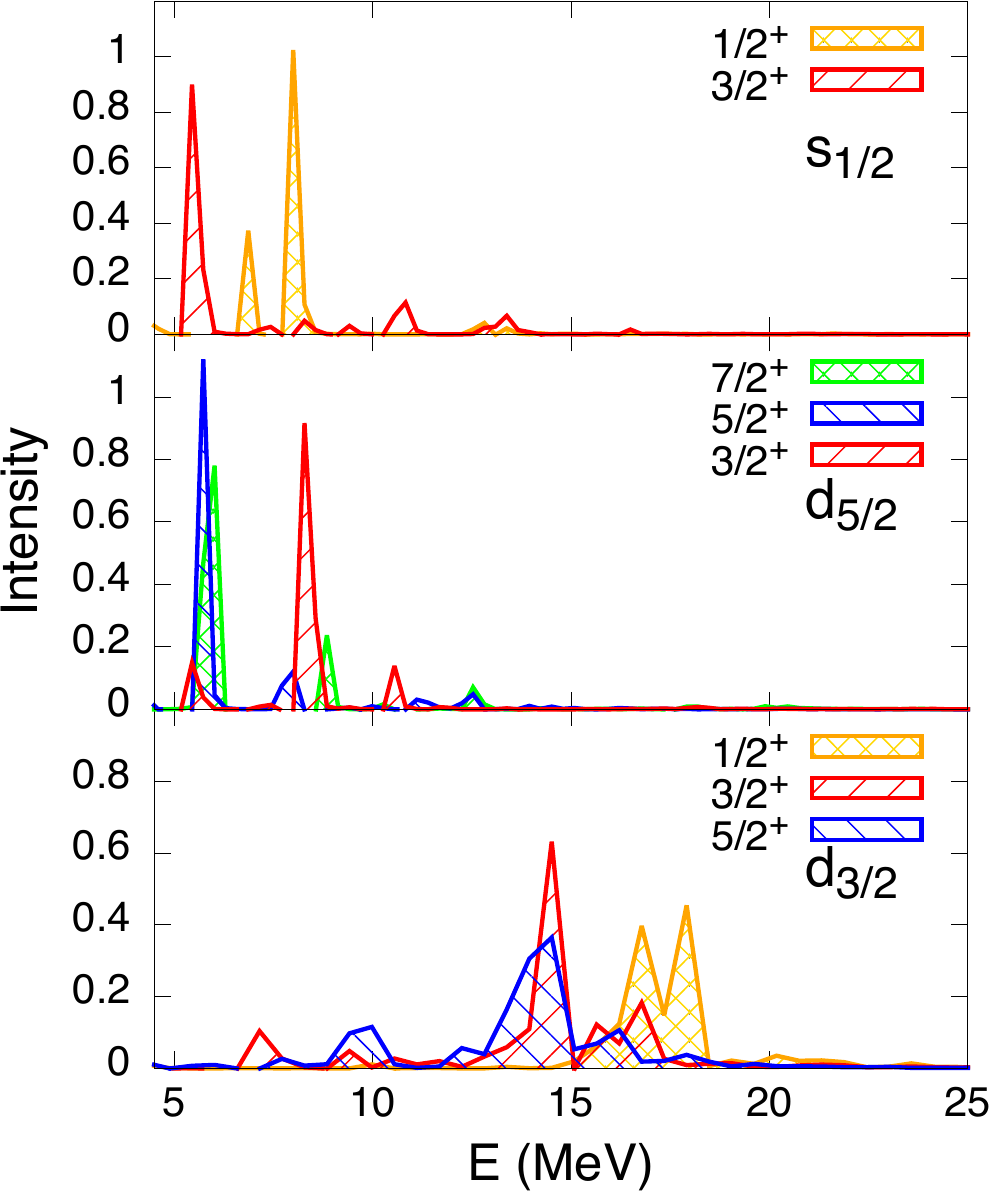}
\caption{(color online) Plot from the theoretical calculations showing where 
the dominant 2s$_{\unitfrac{1}{2}}$, 1d$_{\unitfrac{5}{2}}$, and 
1d$_{\unitfrac{3}{2}}$ states are located in relation to the excitation energy 
in $^{15}$N.}
\label{FIGURE:15N_SF-Theory}
\end{figure}

The experimentally determined excitation energy concentrations of neutron 
single particle strengths for the 2s$_{\unitfrac{1}{2}}$ and 
1d$_{\unitfrac{5}{2}}$ orbits is well reproduced by the shell model calculations 
as displayed in Figure \ref{FIGURE:15N_SF-Theory}. The energy scale begins at 
5 MeV of excitation in $^{15}$N. This figure shows that both the 
2s$_{\unitfrac{1}{2}}$ and 1d$_{\unitfrac{5}{2}}$ strengths lie within the first 
5-10 MeV of excitation and are concentrated in just a few strong levels, 
consistent with the experimental data. In fact the dominant single particle 
strength is quickly exhausted as the nuclear excitation increases so that a 
positive parity state such as the third $\unitfrac{5}{2}^{+}$ state at 9.155 
MeV is almost unobservable in the (d,p) spectrum. In contrast, the 
1d$_{\unitfrac{3}{2}}$ single particle strength is spread out over a region of 
about 4 MeV and is about 5 MeV above the centroid of the lower two orbits. 
In addition levels with the majority of the 1d$_{\unitfrac{3}{2}}$ single 
particle strength are expected to be quite wide since they will be 3$-$5 MeV 
above the $^{14}$N + n separation energy of 10.8 MeV and difficult to identify.

\begin{figure*}[!t]
\centering
\includegraphics[width=2\columnwidth]{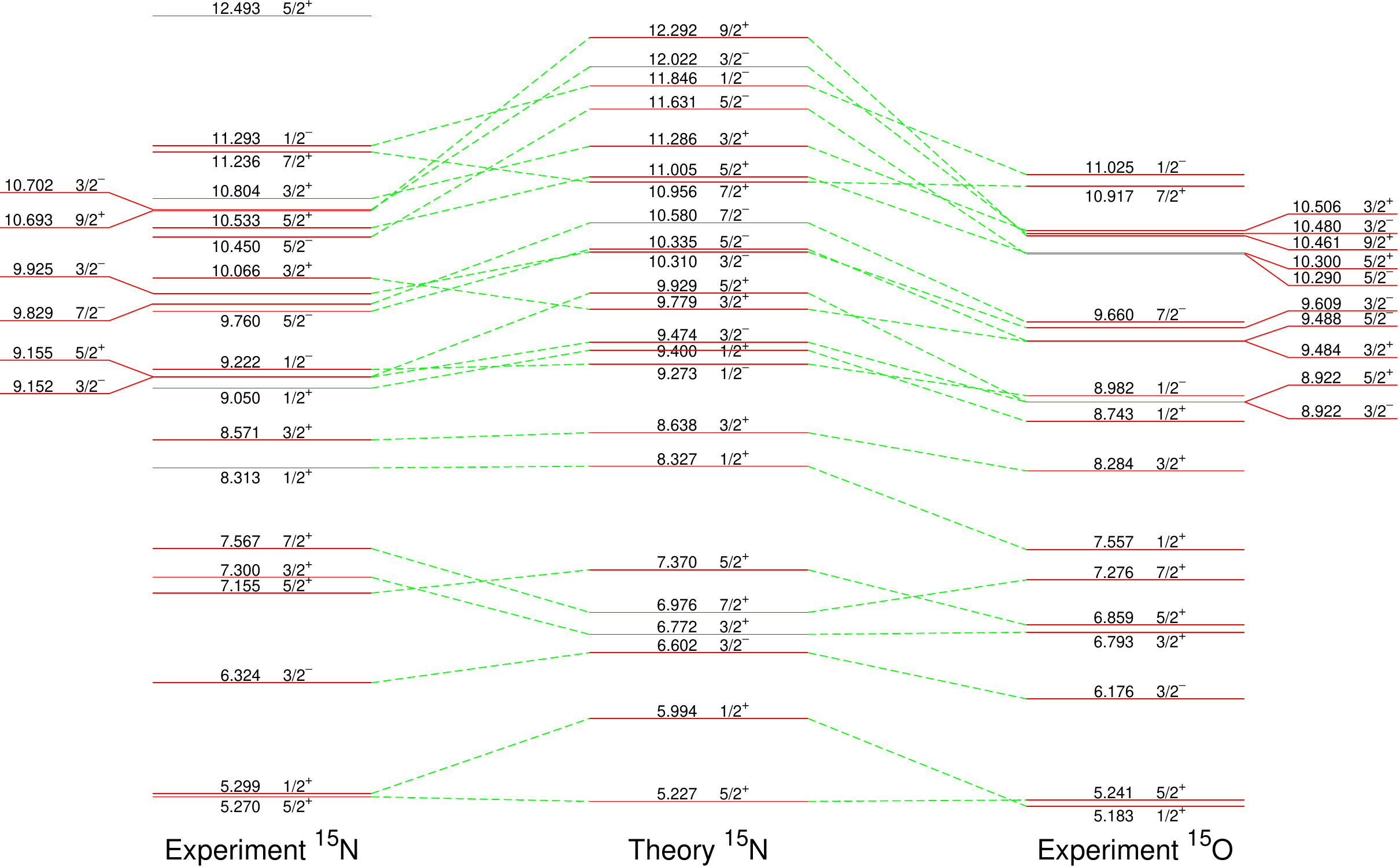}
\caption{(color online) Mirror states between $^{15}$N, both Experimental and 
Theoretical Calculations, and $^{15}$O.}
\label{FIGURE:15N-O}
\end{figure*}

\section{\label{sec:level1}Mirror Levels in $^{15}$N$-^{15}$O}
The structure of the $^{15}$N$-^{15}$O nuclei has been studied both 
experimentally and theoretically for many years now. However, there are still 
levels at relatively low excitation energy ($\sim$less than 12.5 MeV) in 
$^{15}$O whose spin and parities are not determined. There are also levels 
that exist in one or the other of these mirror nuclei but with no 
corresponding partner in the other. Table \ref{TABLE:15N-O-Old} lists levels 
in $^{15}$N and $^{15}$O up to 11 MeV in excitation from the current 
compilation \cite{Ajz91} with possible pairings based on their known 
properties and energy differences. The levels in $^{15}$O are on average 
0.283 MeV below those for $^{15}$N as determined from the known low lying 
levels in both nuclei. The uncertainties in the listed levels for the mirrors 
$^{15}$N$-^{15}$O become readily apparent above about 9 MeV in Table 
\ref{TABLE:15N-O-Old}.

The 9.152 MeV level in $^{15}$N with a well established $\unitfrac{3}{2}^{-}$ 
assignment has no known mirror in $^{15}$O. There are two levels at 8.922 MeV 
in $^{15}$O with one being the $\unitfrac{5}{2}^{+}$ mirror level of the 
$^{15}$N 9.155 MeV level and the other assigned as $\unitfrac{1}{2}^{+}$ in 
$^{15}$O but with no corresponding $^{15}$N mirror. Early reviews 
\cite{Lie76, Amo72} showed this level in $^{15}$O as $\unitfrac{1}{2}^{-}$. 
The reason for the change of  assignment to $\unitfrac{1}{2}^{+}$ is not 
readily apparent from a review of the literature from the time of this change. 
The present mirror level would assign $\unitfrac{3}{2}^{-}$ to the second 
8.922 MeV level with its mirror being the 9.152 MeV $\unitfrac{3}{2}^{-}$ 
level in $^{15}$N. The $^{15}$O level listed at 9.484 
$\left( \unitfrac{3}{2}^{+}\right)$ has no known mirror in $^{15}$N and is 
probably the same as the 9.488 $\unitfrac{5}{2}^{-}$ level with its $^{15}$N 
mirror being at 9.76 MeV. While the current compilation has the spin-parity 
assignments for the $^{15}$O levels at 8.98, 9.829, 10.066, 10.29, 10.461, 
10.48 and 10.506 in brackets, (see Table \ref{TABLE:15N-O-Old}), mirror levels 
in $^{15}$N all have firm assignments which then allows  these brackets to be 
removed. In addition, the 9.66 MeV level in $^{15}$O is assigned as 
$\unitfrac{7}{2}^{-}$ based on its mirror 9.829 in $^{15}$N. The present 
$^{14}$N(d,p) study along with the shell model calculations presented earlier 
assign $\unitfrac{7}{2}^{+}$ to the 11.236 MeV $^{15}$N level. Its mirror at 
10.917 in $^{15}$O has a firm assignment of $\unitfrac{7}{2}^{+}$ supporting 
this assignment. Table \ref{TABLE:15N-15O} gives the final level assignments 
in the mirrors $^{15}$N$-^{15}$O from the present work.

\section{\label{sec:level1}2\MakeLowercase{s}$_{\unitfrac{1}{2}}$ and 
1\MakeLowercase{d}$_{\unitfrac{5}{2}}$ single particle centroid energies in 
$^{15}$N$-^{15}$O}
From the present work and a recent detailed study of the $^{14}$N($^{3}$He,d) 
reaction \cite{Ber02} it is possible to extract the single particle centroid 
energies for the 2s$_{\unitfrac{1}{2}}$ and 1d$_{\unitfrac{5}{2}}$ single particle 
orbits in the $^{15}$N$-^{15}$O pair. The single particle strengths for the two 
reactions are similar in terms of which states have the major single particle 
components but the absolute values for the ($^{3}$He,d) reaction are on average 
only 65\% of those extracted from the (d,p) analysis. Because $^{14}$N has a 
ground state spin of 1, the determination of the energy weighted single 
particle energies (EWSPE) is slightly modified when compared to its 
determination from spin 0 targets and equation 3 of Ref. \cite{Bed13} is used 
to determine the values in the present work. The present $^{14}$N(d,p) work 
showed there to be single neutron strength at higher excitation energies that 
would shift the centroid energies higher from those computed based on previous 
lower energy work. In addition the ($^{3}$He,d) work limited its study to 
lower lying levels so that it is necessary to estimate where high lying single 
particle strength might be found in $^{15}$O if a meaningful comparison 
between the two nuclei is to be made. Data for the $^{15}$O single particle 
strengths were inferred from their $^{15}$N mirror for ``transitions'' to 
$^{15}$O levels at 8.284 $\left(\unitfrac{3}{2}^{+}\right)$, 9.484 
$\left( \unitfrac{3}{2}^{+}\right)$ and 10.917 
$\left(\unitfrac{7}{2}^{+}\right)$ MeV. The spectroscopic factors for these 
``transitions'' were generated from those for the $^{15}$N mirror states by 
multiplying them by 0.6, the factor by which the $^{15}$O spectroscopic 
factors are reduced relative to the $^{15}$N ones. The energy weighted sum 
rule for the 1d$_{\unitfrac{5}{2}}$ orbit in $^{15}$N is then 7.97 MeV while 
that for $^{15}$O is 7.47 MeV. The 2s$_{\unitfrac{1}{2}}$ sum rule is then 8.08 
MeV in $^{15}$N and 7.43 MeV in $^{15}$O. These results show that the 
1d$_{\unitfrac{5}{2}}$ and 2s$_{\unitfrac{1}{2}}$ orbits in these mass 15 mirrors 
are essentially degenerate. If one uses just the main spectroscopic factors 
reported by Bertone {\em et al.} \cite{Ber02}, then the $^{15}$O 
1d$_{\unitfrac{5}{2}}$ centroid is at 6.98 MeV and that for 2s$_{\unitfrac{1}{2}}$ 
is 7.11 MeV and using only their mirror levels in $^{15}$N, the 
1d$_{\unitfrac{5}{2}}$ centroid lies at 7.26 MeV and that for 
2s$_{\unitfrac{1}{2}}$ is at 7.64 MeV.  The higher lying single particle 
strengths, while weak, do raise the single particle centroid energies by about 
0.5 MeV. Based on the location of the 1d$_{\unitfrac{3}{2}}$ orbit in $^{17}$O, 
it is expected that its single particle strength in the mass 15 system will 
lie 5 MeV or so above the 1d$_{\unitfrac{5}{2}}$ orbit and so would be centered 
around 13 MeV in excitation, where a simple calculation shows its width to be 
about 1 MeV, thus confirming the strength calculation shown in Fig. 
\ref{FIGURE:15N_SF-Theory}.

\section{\label{sec:level1}The 11.436 MeV level in $^{15}$N}
An interesting test of the current shell model calculation occurs with the 
prediction of the existence of a $\unitfrac{7}{2}^{-}$ level located at 11.54 
MeV in $^{15}$N. While the current compilation for $^{15}$N does not show such 
a level, a search of the literature reveals just such a level found in a 
$^{11}$B~+~$\alpha$ resonance study by Wang {\em et al.} \cite{Wan91}. Current 
level listings for $^{15}$N show only the presence of a $\unitfrac{1}{2}^{+}$ 
level with a width of 41 keV located at 11.438 MeV whereas the data in Ref. 
\cite{Wan91} show both the presence of this level and a very narrow level 
($\sim 10^{-3}$ keV) which they assign as $\unitfrac{7}{2}^{-}$. Further 
support for this $\unitfrac{7}{2}^{-}$ assignment is found in the 
$^{11}$B($^{7}$Li,t) and $^{11}$B($^{6}$Li,d) work of Norton {\em et al.} 
\cite{Nor76} which populates a peak at 11.44 MeV. Unfortunately the 
experimental energy resolution of the alpha transfer work ($\sim$100 keV) does 
not allow the extraction of experimental widths which would determine whether 
both the $\unitfrac{1}{2}^{+}$ and $\unitfrac{7}{2}^{-}$ levels are populated 
or only one of the pair at 11.44 MeV. However, because there is a one to one 
correspondence between the calculated and experimental levels up to 
$\sim$12 MeV in $^{15}$N it is likely that the new $\unitfrac{7}{2}^{-}$ level 
should be added to the level structure of $^{15}$N.

\section{\label{sec:level1}Conclusion}
The present work reports new $^{14}$N(d,p) data that find no narrow single 
neutron states occurring above about 12.5 MeV in excitation in contrast with 
multi-particle transfer reactions that populate a rich spectrum of states up 
to at least 20 MeV in excitation in $^{15}$N. Shell model calculations that 
employ a cross shell 1p$-$2s1d interaction are able to reproduce and confirm 
the known $^{15}$N positive and negative parity level scheme up to 11.5 MeV in 
excitation and are used to assign the spin-parity to a single particle level 
at 11.236 MeV in $^{15}$N as well as to confirm the presence of a narrow level 
at 11.436 MeV of an alpha particle cluster nature. Knowledge of the 
spin-parities of the $^{15}$N levels is then used to provide a complete set of 
spin-parities for its less well known mirror, $^{15}$O, up to 11 MeV in 
excitation. Figure 7 displays the calculated and final level schemes for 
these nuclei. The 2s$_{\unitfrac{1}{2}}$ and 1d$_{\unitfrac{5}{2}}$ single 
particle centroid neutron and proton energies show these orbits to be 
degenerate in the mass 15 system which means that these are the nuclei where 
the level ordering shifts between the lighter systems where the 
2s$_{\unitfrac{1}{2}}$ orbit lies below that of the 1d$_{\unitfrac{5}{2}}$ orbit 
and the heavier system where the 1d$_{\unitfrac{5}{2}}$ orbit lies below that of 
the 2s$_{\unitfrac{1}{2}}$ orbit.

\acknowledgements{This work was supported in part by the U.S. National Science 
Foundation.}

\end{document}